# Квантовая криптография в аспекте популяризации науки и развития профессионально-технической квалификации

Л.И. Стефаненко[1], А.Г. Сергеев[2], Ю.В. Курочкин[2,3,4],

В.Е Родимин[2,3,4]

Научный руководитель: к.ф.-м.н. Родимин Вадим Евгеньевич

[1]Институт новых материалов и новых технологий, Национальный исследовательский технологический университет «МИСиС», Москва, Ленинский проспект, д. 4, 119049, Россия,

[2]Российский квантовый центр, Сколково, Москва 143025, Россия,

[3]Центр квантовых коммуникаций НТИ, Национальный исследовательский технологический университет «МИСиС», Ленинский проспект, д. 4, 119049 Москва,

[4]ООО «КуРэйт», Сколково, Москва 143025, Россия

lis@rqc.ru

## Аннотация

Ореол загадочности вокруг квантовой физики затрудняет продвижение квантовых технологий. Для демистификации нужны методические приемы, объясняющие основы квантовых технологий без метафор и абстрактной математики. Статья дает пример такого объяснения для протокола квантового распределения ключа BB84 на базе фазового кодирования. Это позволяет органично перейти к знакомству с реальной криптографической установкой QRate, используемой на конкурсе WorldSkills в компетенции «Квантовые технологии».

## Ключевые слова

Оптика, квантовая оптика, образование, популяризация науки, квантовая криптография, квантовая физика, квантовые технологии, WorldSkills, Future Skills, интерферометр Маха–Цендера, протокол BB84, протоколы шифрования

**Введение.** С момента появления квантовой механики прошло уже около 100 лет. Мы каждый день пользуемся техническими устройствами, основанными на ее достижениях, такими как лазеры и полупроводниковые приборы. При этом уровень понимания даже самых основных идей квантовой механики вне физического сообщества остается чрезвычайно низким. Он уступает даже уровню понимания основ теории относительности. Хотя у релятивистской физики почти нет повседневных применений (одно из немногих исключений — спутниковые навигационные системы), большинство образованных людей имеет какое-то представление о замедлении времени при больших скоростях и об искривлении пространства вблизи массивных объектов. В то же время образы, связанные с квантовой механикой, обычно не идут дальше пресловутого кота Шредингера, который «одновременно жив и мертв».

Недостаток понимания идей квантовой физики приводит к тому, что само слово «квантовый» начинает восприниматься как синоним чего-то загадочного, непостижимого, а то и сомнительного. Возникающие из-за этого ошибки порой приводят к неадекватному медийному отражению развития квантовых технологий, как это было с невежественной реакцией прессы в июне 2016 года на заявление российских представителей власти о перспективах квантовой телепортации, которую некоторые СМИ преподнесли как телепортацию из научной фантастики. Такие недоразумения могут мешать формированию необходимого доверия к квантовым технологиям со стороны бизнес-сообщества, политиков и публики.

Отчасти причиной подобного положения дел может быть недостаточное внимание самих физиков к разработке научно-популярного и педагогического дискурса, позволяющего корректно вводить идеи квантовой механики в контекст общей культуры. Десятилетиями физики бравировали своего рода эзотеричностью квантовой механики. Журналисты любят цитировать классиков квантовой физики: «Тот, кто не шокирован квантовой теорией, просто ее не понимает» (Нильс Бор); «Можно смело сказать, что квантовую механику не понимает никто» (Ричард Фейнман); «Квантовая механика абсолютно лишена смысла» (Роджер Пенроуз) [1].

Действительно, для многих явлений квантового масштаба пока не найдено аналогов в обыденном мире, и их трудно выразить нашим повседневным языком. В результате введение в квантовую механику начинается с описания математического формализма, что становится препятствием для неподготовленного слушателя и не годится для популяризации. Более того, такой подход создает проблемы и для самих физиков: многие из них, изучая математический формализм квантовой механики и убеждаясь на практике в его эффективности, не задаются вопросом о том, как основатели этой науки смогли пройти в обратную сторону и, отталкиваясь от опыта, найти нужный формализм. В результате даже из поля зрения профессиональных физиков могут выпадать важные проблемные и эвристические моменты развития их науки.

С учетом сказанного важной задачей является поиск эффективных объяснительных приемов, позволяющих рассказывать о квантовых явлениях без обращения к математическому аппарату и без злоупотребления сомнительными метафорами. В настоящей статье делается попытка найти такой подход к объяснению квантовой криптографии. Выбор в ее пользу определяется тем, что это одна из самых зрелых квантовых технологий второго поколения, которая уже начинает порождать спрос на технических специалистов для наладки и обслуживания линий защищенной квантовой связи. Для этой деятельности необходимо общее понимание идей квантовой механики, лежащих в основе технологии, но в то же время не требуется полное владение математическим аппаратом, который используется в исследованиях и разработках. В соответствии с этим была поставлена задача изложить принцип квантовой криптографии без обращения к таким абстрактным концепциям, как пространство состояний квантовой системы, его базисы и выбор между ними при квантовом измерении.

В научно-популярном смысле преимущество квантовой криптографии в том, что она достаточно проста для понимания. Наш опыт показывает, что ее основные положения можно объяснить студентам примерно за полчаса без экстраординарных умственных усилий. Также она загадочна и увлекательна, поскольку, с одной стороны ассоциируется со шпионскими шифрами, а с другой — обеспечивает защиту, основанную на законах природы. В методическом плане ценен позитивный аспект квантовой криптографии. Многие фундаментальные положения квантовой физики носят негативный характер: невозможность провести любое измерение без воздействия на систему, принцип неопределенности Гейзенберга, теорема о запрете клонирования. При этом квантовая криптография напротив *разрешает* абсолютно безопасное распределение ключа шифрования.

Мы предлагаем способ научно-популярного объяснения ряда идей квантовой физики на примере реализации протокола BB84

использованием фазового кодирования передаваемой информации. На базе этого объяснения можно органично переходить к знакомству с реальной установкой квантового шифрования, созданной Российским квантовым центром и компанией QRate [2]. В свою очередь эта установка стала базой для формирования и развития новой компетенции «Квантовые технологии» в рамках всемирной системы профессиональных конкурсов WorldSkills.

**Материалы и методы исследования.** Интерферометр Маха–Цендера (ИМЦ) состоит из двух светоделителей — полупрозрачных зеркал — и двух обычных зеркал (см. рис. 1) Входящий луч делится пополам, разводится по разным плечам, затем лучи сводятся вновь на втором светоделителе-зеркале и расходятся на детекторы. Если пути, которые проходят два луча, одинаковые, то такой интерферометр называется равноплечим.

На первый взгляд, лучи на выходе равноплечего ИМЦ должны иметь одинаковую интенсивность, равную половине интенсивности исходного луча. Так и происходит, если запускать в ИМЦ обычный некогерентный (например, естественный) свет и измерять освещенность на двух выходах. Однако при использовании когерентного лазерного излучения результат оказывается существенно иным. Это связано с волновой природой света и особенностями работы светоделителей, а именно с тем, что отраженная и преломленная световые волны по-разному сдвигаются по фазе. Не умаляя общности, можно считать, что между отраженной волной и той, которая прошла светоделитель насквозь, возникает фазовый сдвиг на четверть длины волны ($\lambda/4$) или, в терминах фазы, на $\pi/2$ [3].

Луч, который выходит из ИМЦ на $D1$, формируется двумя составляющими, испытавшими по одному отражению на светоделителях. В равноплечем интерферометре они имеют одну и ту же фазу и демонстрируют конструктивную интерференцию. Луч же, выходящий из ИМЦ на $D2$, складывается из двух составляющих, испытавших, соответственно, ноль и два отражения на светоделителях. Это дает разность хода $\lambda/2$ и фазовый сдвиг $\pi$, то есть, составляющие находятся в противофазе и испытывают деструктивную интерференцию. В результате для когерентного входящего луча равноплечий ИМЦ формирует только один исходящий луч на $D1$, равный по интенсивности исходному

(потерями пренебрегаем). На другом выходе *D*2 световые колебания полностью гасятся.

Если теперь в одно из плеч интерферометра добавить фазовый модулятор (*PM*), вызывающий сдвиг фазы электромагнитной волны на полпериода ($\pi$), то ситуация изменится на противоположную. Теперь на *D*1 интерференция будет деструктивной, а на *D*2 — конструктивной.

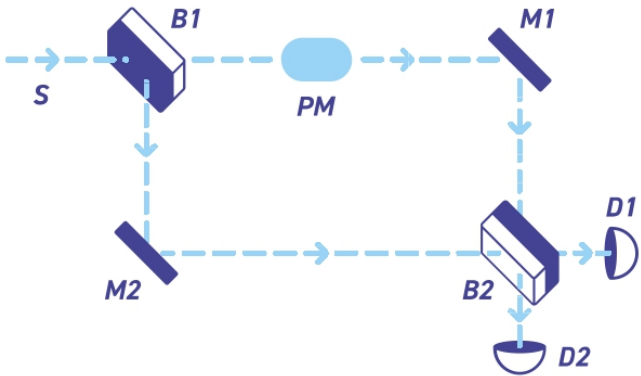

Рис. 1. Схема интерферометра Маха–Цендера. *S* — источник света, *PM* — фазовый модулятор, *B*1, *B*2 — полупрозрачные зеркала, *M*1, *M*2 — обыкновенные зеркала, *D*1, *D*2 — детекторы одиночных фотонов

Включая и выключая такой фазовый модулятор, можно переключать выход луча с *D*1 на *D*2. Если с помощью фазового модулятора сделать сдвиг $\pi/2$ или $3\pi/2$, то свет будет попадать на оба детектора *D*1 и *D*2, на каждый детектор интенсивность приходящего света будет равна половине интенсивности входящего.

**Однофотонный режим.** Теперь от классического рассмотрения работы ИМЦ перейдем к квантовому приближению. Будем ослаблять интенсивность входящего света до тех пор, пока мы не выйдем на уровень дискретизации света — на квантовый уровень. Для этого детекторы *D*1 и *D*2 должны быть достаточно чувствительными, чтобы регистрировать одиночные фотоны. Слабый свет от лазера будет регистрироваться детекторами в виде одиночных срабатываний. Эти срабатывания по времени будут носить случайный, хаотический характер. При этом в среднем количество срабатываний в единицу времени будет соответствовать интенсивности света.

Как было показано выше, в случае классического, волнового рассмотрения света, в равноплечем интерферометре на $D1$ происходит конструктивная интерференция. Согласно принципу соответствия [4], переходя на квантовый уровень по мере ослабления света, переход должен происходить естественно и плавно в плане наблюдаемых явлений. Поэтому и на однофотонном уровне конструктивная интерференция будет на $D1$, т.е. все фотоны будут приходить на первый детектор.

Концептуальная проблема понимания происходящего возникает, если вдуматься, как получается, что одиночный фотон приходит на первый детектор. Ведь именно так следует из эксперимента. Более того, вместо лазера можно использовать источник одиночных фотонов, который используется в лабораториях, и контролируемо посылать в ИМЦ только один фотон, всякий раз убеждаясь, что сработал $D1$. Отдельный фотон не может разделиться на два. Иначе разделилась бы его энергия, и частота излучения снизилась вдвое. Но нельзя и предположить, что каждый отдельный фотон случайным образом проходит через ИМЦ по какому-то одному пути — верхнему или нижнему. Тогда фотонам было бы не с чем интерферировать на выходе, и они бы с равной вероятностью регистрировались обоими детекторами, а это противоречит эксперименту.

Любая попытка зафиксировать прохождение фотона по одному из путей приводит к тому, что он попадает на второй светоделитель с известного направления и далее случайным образом перенаправляется в один из детекторов. Причем такой результат имеет место даже тогда, когда предпринятая попытка зарегистрировать фотон в одном из плеч интерферометра оказалась неудачной, поскольку в этом случае известно, что фотон прошел по другому плечу.

Таким образом, одиночный неотслеживаемый фотон и не разделяется, и не выбирает один из двух путей через ИМЦ, а каким-то образом проходит через интерферометр сразу по двум путям, оставаясь при этом единой частицей. О таком фотоне говорят, что он находится в состоянии квантовой суперпозиции фотона, идущего по верхнему пути, и фотона, идущего по нижнему пути. Квантовая суперпозиция — это математическая абстракция, лежащая в основе квантовой физики. Достоверно неизвестно, какие свойства материального мира приводят к квантовой суперпозиции.

Каждая составляющая этой суперпозиции математически описывается аналогично максвелловской волне и имеет ту же частоту. И хотя в данном случае нельзя говорить о колебаниях вектора электрического поля, тем не менее, такие волны также испытывают влияние фазовых модуляторов и интерферируют друг с другом. Когда составляющие суперпозиции после второго светоделителя ($B2$) складываются, они гасят или усиливают друг друга на детекторах соответственно разности своих фаз.

В равноплечем ИМЦ разность фаз по двум путям равна нулю. Поэтому при интерференции составляющих суперпозиции одиночного фотона получается тот же результат, что и в случае классической электромагнитной волны: конструктивная интерференция на детекторе $D1$, который щелкает на каждый фотон, и деструктивная интерференция на $D2$, который, соответственно, молчит.

Добавление в плечи ИМЦ оптических элементов, меняющих фазу волны, влияет на интерференцию фотона с самим собой на выходе. В частности, при относительном сдвиге фаз на $\pi$, реакция детекторов меняется на противоположную — $D2$ щелкает, а $D1$ молчит.

**Результаты и обсуждения.** Рассмотрим задачу конфиденциальной передачи информации из пункта А в пункт Б. С 70-х годов прошлого столетия такой приемник и передатчик принято именовать Алисой и Бобом, соответственно. Пусть плечи ИМЦ будут длинными, так что с одной стороны будет находиться Алиса, а с другой стороны — Боб. Эти персонажи могут использовать ИМЦ для передачи двоичной информации на однофотонном уровне. Алиса включает фазовый модулятор, когда передает «1», и выключает его для передачи «0». А у Боба меняются условия интерференции и срабатывает, соответственно, детектор $D1$ или $D2$. Несмотря на однофотонный режим, такой информационный канал не защищен от прослушивания. Злоумышленник, традиционно именуемый Евой, может проникнуть в такой разнесенный интерферометр, провести измерение вместо Боба, и отправить Бобу результат измерения в виде фотона в суперпозиции прохождения по двум плечам, не отличимый от того, как это делает Алиса. Иначе говоря, вместо одного ИМЦ, соединяющего Алису и

Боба, будет работать два ИМЦ, соединяющих Алису и Еву, а затем — Еву и Боба.

Чтобы получить секретный канал передачи информации, процедуру приготовления и измерения фотона надо усложнить. Это сделано в протоколе BB84, разработанном еще в 1984 году [5]. Пусть, как и выше, $D1$ отвечает за бит «0», $D2$ – за бит «1». Пусть теперь Алиса будет прикладывать фазовые сдвиги $\pi/2$ и $3\pi/2$. Если при этом Боб на своем модуляторе приложит сдвиг $\pi/2$, то они опять смогут передавать информацию на однофотонном уровне. Когда Алиса шифрует бит, прикладывая 0 или $\pi$, а Боб проводит измерение, не делая фазового сдвига, то говорят, что Алиса готовит фотон в базисе I, а Боб измеряет фотон в базисе I. Когда Алиса шифрует бит, прикладывая $\pi/2$ или $3\pi/2$, а Боб проводит измерение, прикладывая фазовый сдвиг $\pi/2$, то говорят, что Алиса готовит фотон в базисе II, а Боб измеряет фотон в базисе II. Если Алиса и Боб работают в одинаковом базисе, то они могут передавать информацию; если они работают в разных базисах, то у Боба будет срабатывать случайный детектор. В итоге они затевают такую игру:

- Алиса случайным образом выбирает базис приготовления фотона, значение бита, готовит фотон и пересылает Бобу
- Боб выбирает случайный базис измерения фотона, проводит измерение и записывает получившееся значение бита
- В половине случаев он случайно угадал базис приготовления фотона, тогда у него получился тот бит, который ему послала Алиса. В другой половине случаев результат Боба получился случайным
- По открытому каналу Алиса и Боб сообщают друг другу, какие базисы приготовления и измерения они использовали. Но держат в секрете приготовленные и измеренные значения бит!
- Алиса и Боб выбрасывают те значения, которые соответствуют разным базисам. В идеальном случае оставшиеся битовые последовательности Алисы и Боба будут одинаковые.

Это самый первый и самый распространенный протокол квантовой криптографии — квантового распределения ключа. Можно сказать, что квантовое распределение ключа — это такой способ генерации случайной битовой последовательности (ключа) в двух удаленных друг от друга местах. В обоих местах ключ получается одинаковым, и при этом секретным.

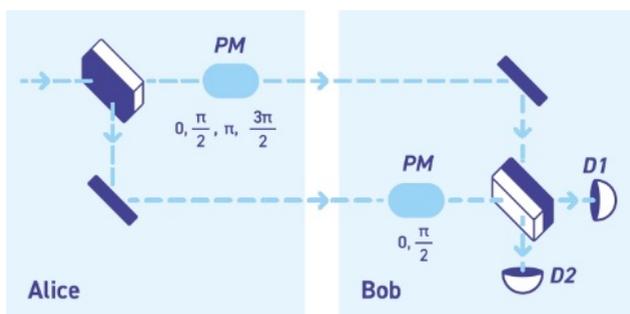

Рис. 2. Схема разнесенного интерферометра Маха–Цендера. *PM* — фазовый модулятор, *D*1, *D*2 — детекторы одиночных фотонов

Если Ева захочет вмешаться, она немного сможет сделать. Ей придется случайным образом решать, какой базис измерения выбирать. В результате она начнет генерировать ошибки в ключах, рандомизированными окажутся три четверти переданных битов, Алиса и Боб при сверке не смогут отсеять «испорченные» биты и в итоге получат разные ключи. Достаточно сверить их фрагменты или контрольные суммы, чтобы обнаружить постороннее вмешательство и прекратить передачу данных до его устранения.

Осмысленной информации ключ не несет, т.к. это случайная битовая последовательность. Но Алиса и Боб могут использовать свои ключи для шифрования и дешифровки сообщений. Квантовая криптография дает возможность осуществлять безусловно секретную передачу данных, секретность такого шифрования доказывается строго математически.

**Заключение.**

**Компетенция «Квантовые технологии» на WorldSkills.** Сегодня большая часть интернет-трафика передается по оптическим линиям связи. Поэтому у крупных компаний в штате или на аутсорсе есть специалисты, умеющие настраивать такие линии. Ожидаемый рост спроса на линии связи, защищенные квантовым шифрованием, создаст потребность в специалистах, умеющих работать с этой квантовой технологией. Это означает, что формировать соответствующую компетенцию и отрабатывать подготовку и проверку квалификации будущих специалистов необходимо уже сейчас.

С пониманием этой перспективы РКЦ и QRate присоединились к международному проекту WorldSkills, целью которого является повышение статуса различных профессий и развитие стандартов профессиональной подготовки и квалификации. Достигается это через организацию конкурсов профессионального мастерства на региональном и международном уровне.

В 2017 году в рамках проекта Future Skills WorldSkills Russia была учреждена новая компетенция «Квантовые технологии». На сегодня значительная часть навыков, по которым проводится соревнование в этой компетенции, относится к налаживанию квантово-защищенного канала связи на базе академической установки квантового распределения ключа QRate.

В ходе соревнований участники должны выполнить полный спектр задач по запуску стабильного распределения квантового ключа с помощью одиночных фотонов: создать оптоволоконный канал, собрать оптические схемы Алисы и Боба, которые будут обслуживать квантовый канал, провести калибровку квантово-оптической линии и запустить процесс распределения квантового ключа, а также презентовать результаты настройки системы.

Дополнительно выполняются задания по исследованию характеристик однофотонных детекторов, которые используются в системе, а также по реализации алгоритма, выполняемого на свободно доступном для публичных экспериментов квантовом процессоре IBM. Последнее задание включено в программу соревнований с тем, чтобы квалификация «Квантовые технологии» не сводилась к одной только квантовой коммуникации, а также с целью популяризации идеи квантовых вычислений и квантовых компьютеров.

С 2017 по 2021 год было проведено 24 чемпионата, включая соревнования на Чемпионате мира в Казани в 2019 году, выступления на национальном финале и представление компетенции на международном уровне в Китае. Компетенция включена во все линейки чемпионатов: региональные, вузовские, корпоративные, национальные, международные.

Кроме того, специалистами компетенции был создан комплект оценочной документации для проведения Демонстрационного экзамена, проведена сертификация экспертов, разработаны методические рекомендации и комплект конкурсной документации.

Проделанная работа по развитию Компетенции позволила сформировать специальность 11.02.13 Твердотельная электроника (Лот №3 Информационные и коммуникационные технологии) в перечне профессий и специальностей среднего профессионального образования.